# The study of $Ag^+(Cu^+)$ center creation in quartz crystals by thermal annealing at $1200^oC$ in $O_2$ atmosphere from metallic Ag and Au

Anatoly Trukhin, Madara Leimane

Institute of Solid State Physics, University of Latvia, Kengaraga St.8, LV-1063, Riga, Latvija

## Abstract

The study of $Ag^+(Cu^+)$ center creation in α-quartz crystals by thermal annealing at $1200^oC$ in $O_2$ atmosphere from metallic Ag and Au. Previously, the effect of treatment in ($O_2$) atmosphere at high temperatures on copper (Cu) luminescence centers in α-quartz was discovered [1]. In this work, we investigated the effect of such treatment on silver (Ag) centers in α-quartz, studied previously [2]. This treatment also affects the luminescence center of Ag, but not as effectively as in the case of Cu. We believe that the effect is determined by the duration, which previously was 100 hours, but now we can process no more than 2 hours. Also, a series of non-activated crystalline α-quartz samples were thermally annealed at $1200^°C$ in an $O_2$ atmosphere together with metallic Ag.

A characteristic $Ag^+$ luminescence center with a band at 260 nm was discovered for α-quartz samples thermally annealed in $O_2$ and initially not containing Ag. It is likely that at high temperatures an oxygen-silver complex is formed, which can diffuse into previously non-activated samples, and modification of the existing aluminum-alkali metal center occurs, probably due to the replacement of the alkali ion ($Me^+$) by $Ag^+$.

An attempt was made to introduce old (Au) into α-quartz by such treatment in $O_2$. Both high-purity Au and Au-containing parasitic impurities of Ag and Cu were used. In the latter case, we obtained $Ag^+$ and $Cu^+$ luminescence centers but were unable to identify the luminescence of centers associated with Au. In the case of using high-purity Au, neither centers associated with Cu and Ag nor centers that could be associated with Au were found. The luminescence center associated with Au was not identified using pure Au as a precursor source and when attempting to introduce Au by electrolysis.

## Introduction

It was previously established that crystalline quartz containing $Cu^+$ impurities [1] and annealed at a temperature of 1200°C in an $O_2$ atmosphere is modified, and a complex impurity center containing aluminum (Al) and a charge compensator $[AlO_4^-–Me^+]$ ($Me^+$ in this case is the $Cu^+$ ion) ([2] and references therein) changed their luminescent properties. Therefore, it can be assumed that treating silicon dioxide ($SiO_2$) in $O_2$ atmosphere at high temperatures may be a method for modifying the impurity defect. To test this idea, this paper examined the luminescence of crystalline quartz samples, both intentionally non-doped and doped with Ag by high-temperature electrolysis. All samples were processed in a flow of $O_2$ at a temperature of 1200°C. A broad overview of the luminescent properties of quartz crystal in particular and $SiO_2$ in general can be found in [2].

**Experimental**

The present samples were various types of crystalline quartz. Samples of natural quartz were smoky and morion, containing impurities of Al and alkali metal ions. In addition, synthetic crystalline quartz containing germanium (Ge) or phosphorus (P) has been studied. Among the all samples nominally pure synthetic quartz was used as a reference sample. The Ag-doped sample was a natural smoky quartz crystal subjected to electrolysis at a temperature of 800°C in Ag plate electrodes with an applied voltage of about 3 kV, as was done previously in [3]. Electrode marks remained on the sample and that was later found to be a source of Ag. A similar experiment was made by introducing Au wire with unknown purity for incorporation of Au in quartz crystal samples. An additional experiment was performed with high-purity Au to study the effect of precursor purity. The dimensions of the studied samples were about 12x12x2 $mm^3$ or smaller.

All the samples were placed in a quartz tube, which was located in the high temperature tube furnace system. The sample annealing process was carried out for 3-4 hours in a pure $O_2$ flow atmosphere and the maximum temperature of 1200°C was reached. The temperature ramp around the α-β transition (~573°C) of quartz was kept low (~1°C/min) to prevent the destruction of the quartz crystal during the thermal annealing process. Based on the positive result with Ag doping in crystalline quartz, we also carried out the same thermal annealing process in $O_2$ atmosphere using metallic Au wire as a metal source. However, Au wire also contained noticeable impurities of Cu andAg.

The luminescence spectra and luminescence decay kinetics were measured. Luminescence was excited either by X-ray (anti-cathode W 40 kV, 20 mA) or ArF (193 nm) or $F_2$ (157) nm excimer lasers. Luminescence detection was carried out either by a Hamamatsu C10082CAH CCD mini-spectrometer (not very sensitive, but useful for obtaining express spectra), or by a MCD grating monochromator with the Hamamatsu R11715-01 photomultiplier. The photon counting module H8259-02 was also used, where pulses transformation was made using an integrator based on MOSFET transistor, and the signal of which was recorded. To determine the steady-state intensity, an Agilent 34411A voltmeter (with a 100 KΩ resistor) was used. Picoscope 2208A (with a 50 Ohm resistor) was used to measure the luminescence decay kinetics, and fast PMT H6780-04 was used for search of fast decay as well. The sample temperature was kept from 15 K to 700 K on different sample holders. The photoluminescence excitation spectra was measured using 0.5 m 70 degree vacuum monochromator Seya-Namioka type and windowless hydrogen flow, RF alimentation about 25 W, discharge light source.

## Results and discussion

*Thermal annealing at 1200 °C in $O_2$ atmosphere of Ag-doped and non-doped α-quartz samples.*
Figure 1 shows the photoluminescence (PL) and photoluminescence excitation (PLE) spectra of quartz doped with Ag by high-temperature electrolysis. One of the sample was cutted into two pieces. One piece was subjected to thermal annealing at 1200 °C in $O_2$ atmosphere and the other piece was left as reference sample. The PLE spectrum of quartz crystal sample studied previously [1] is presented for comparison. For the last sample, which was treated in $O_2$ at 1200°C for 100 hours, significant changes in PLE spectra were obtained and those are: the PLE band at 5.4 eV is practically not observed and the band just below intrinsic absorption of quartz (8.5 eV) is strongly diminished. Whereas the bands related to energy transfer from matrix to $Cu^+$ luminescence center becomes more expressed as in non-treated samples. The actual experiment with Ag-doped quartz the changes, however observed, are not so expressed and are vice versa to the case of Cu. Perhaps that is related due to much shorter period of thermal treatment at 1200°C, but we are not able perform thermal treatment for more than 2 hours.

PL spectra of natural α-quartz heat-treated in $O_2$ at 1200 °C in the presence of metallic Ag is showed in Figure 2. Metallic Ag occasionally remained on the sample previously subjected to high temperature electrolysis. PLE spectra was done by $F_2$ excimer laser (157 nm), PL spectra of Ag-doped α-quartz using electrolysis process is showed for comparison. Its intensity is devised by 10, so underlining that the quantity of Ag that appeared due to treatment is negligible. In addition, it is observed that the band at 360 nm diminishes for Ag-doped α-quartz using the electrolysis process, whereas the band at 360 nm is dominant for the sample heat-treted at 1200°C in $O_2$ atmosphere. It is worth pointing out that $Cu^+$ in α-quartz and $Ag^+$ in silica glass are situated in the same place. However, they possess different PL decay kinetics [2]) belonging to [AlO$_4^-$-Na$^+$(or Li or K)], and during the electrolysis process alkali ions are replaced with Ag ions. During our thermal annealing experiments at 1200 °C in O$_2$ flow atmosphere, the replacement of these ions takes place only partly. In the Figure 3, analogous to Figure 2, the PL spectra was obtained using X-ray as excitation source. The source of Ag is presumably the remains of metallic Ag on the surface of the sample after thermal treatment with high-temperature electrolysis from metallic Ag electrodes.*Treatment in O$_2$ of non-doped α-quartz samples in the presence of low-probe Au.*

In figure 4, the PL spectra of natural quartz heat-treated in O$_2$ at 1200°C in the presence of metallic Au (low-probe) is showed. Figure 4 shows PL decay kinetics parameter of natural crystalline quartz kept at 1200°C in O$_2$ atmpsphere in the presence of a Au wire. The PL spectra of Au-doped sample were measured using PMT through a grating monochromator. In this case, we observed two bands at ambient temperature. One is located at ~260 nm and another at ~360 nm. The PL bands are due to Cu$^+$ (~360 nm with time constant about 50 μs) and Ag$^+$ (260 nm with time constant about 25 μs). In the left part of spectra, the band at 193 nm is from the sample holder excitation laser. The band at 260 nm decays (157 nm excitation with F$_2$ laser) with a time constant of 23 μs, indicating that this is due to Ag$^+$ [2]. For the band at 360 nm, the decay kinetics was measured with three excimer lasers: KrF (248 nm), ArF (193 nm), and F$_2$ (157 nm). In all cases, decay exhibits curves providing time constant ~40 μs, and it is due to Cu$^+$ luminescence center [2]. Cu as well as Ag are the main impurities found in Au wire used in our experiments.The PL spectra of the sample described on the previous figure was measured with PMT through a grating monochromator. In this case, we can observe two bands at ambient temperature. One is at ~260 nm and another at ~360 nm. In the left part of spectra the band at 193 nm comes from the the sample holder excitation laser light. The PL band at 260 nm decays (157 nm excitation with F$_2$ laser) with a time constant of 23 μs, indicating that this is due to Ag$^+$. The other PL band at 360 nm is measured with three excimer lasers (KrF – 248 nm, ArF-193 nm and F2 – 157 nm). In all cases, PL decay kinetics exhibit curves providing time constant ~40 μs, and it is certainly due to Cu$^+$ center. Cu as well as Ag are the main impurities in used gold wire.

Figure 5 shows PL spectra, decay kinetics, and temperature influence on PL parameter of natural crystalline quartz heat-treated at 1200°C in O$_2$ in the presence of Au wire, are showed. The PL spectra was performed with minispectrometer Hamamatsu (main part of the figure). The luminescence decay kinetics were measured with fast Hamamatsu PMT (0.7 ns rising time) together with UFS1 filter, which is transparent in the range of 260-400 nm (upper inset), and using ArF excimer laser (193 nm). The waves in the decay curves are due to some properties of fast PMT (not observed with other PMT). The inset in lower right (in Figure 5) are time resolved temperature dependences of luminescence intensities (integrated decay curves for each temperature) as well as time constant thermal dependencies. The luminescence band at 300-500 nm is composed and contains a subband at about 360 nm at ambient temperatures ~280 K, and increases with cooling blue band at 420 nm. Non-selective decay measurements show two main decay components. One with a duration of about 20 μs, performing thermal quenching at temperatures above 500 K. This is observed for both the decay time constant and the time-resolved intensity. Another slower decay occurs above 200 K. The inset dot shows the temperature dependence of the decay time constant of the slow component. Measuring the decay kinetics of the luminescence band through selective filters is shown in the following figure. In the way to study influence of $O_2$ treatment at high temperature (1200°C) we had checked many samples of quartz crystal among them one which was doped with Ag using high temperature electrolysis. On the last sample, some quantity of Ag electrode remains and as result of $O_2$ treatment, a Ag related PL band appeared practically in all heat-treated samples. The intensity of Ag related PL is correlating with the concentration of impurity center possessing alkali ions as charge compensator, therefore in nominally pure sample the induced Ag PL intensity is negligible. Figure 5 presents all spectra, however details of performed experiment are considered below.

*Treatment in oxygen of non-doped α-quartz samples in the presence of high-probe gold*

Figure 6 shows, that in all natural quartz samples heat-treated at 1200°C in $O_2$ flow atmosphere in the presence of pure Au, band of $Cu^+$ luminescence center was observed. The similar PL characteristics was observed for quartz crystal samples deliberately activated with Cu using electrolysis (KP-Cu($O_2$) and KP-Cu). The possible Cu cross-contamination is due to the Cu impurities in silica quartz tube, where heat-treatment process was performed. This is evident that the experiments in $O_2$ flow atmosphere promote the diffusion of impurities from the reactor to the analyzing samples. In many different natural quartz samples heat-treated with pure Au in $O_2$ at 1200°C, observed unequivocally the band of $Cu^+$ center, the same as quartz crystal samples deliberately activated with Cu by electrolysis (Q-Cu($O_2$)) and Q-Cu). The possible Cu cross-contamination is due to the Cu impurities in silica quartz tube.

*Try to introduce gold in natural quartz sample by electrolysis using high-probe gold as a negative electrode.* The electrolysis experiment was made at 850°C applying 2 kV bias.

Same from the previous experiments, samples for Au-doping experiments were placed in a quartz tube, which was located in the high temperature tube furnace system. The sample annealing process was performed at 850°C. The temperature ramp around the α-β transition (~573°C) of quartz was kept low (~1°C/min) to prevent the destruction of the quartz crystal during the thermal heating with electrolysis process. The initial current for electrolysis process was about 50 µA falling to less than 1 µA during 40 minutes. However, applied bias was kept on the sample till cooling to ambient temperatures. After performing the electrolysis process for Au incorporation in natural quartz sample, the PL spectra was measured and is showed in Fig.7. The PL spectra and PL decay show [$AlO_4^-$-(alkali)$^+$] luminescence center without any significant changes [2]. *Introducing of Cu, Ag, and Au (pure) by high-temperature electrolysis into silica glass containing sodium.*

Initially, we performed experiments by trying to introduce metals (Cu, Ag, and Au) in silica glass (with 0.025 wt% Na) using thermal-annealing process in $O_2$ flow atmosphere in the presence of a piece of pure Au. However, at that temperature, the sample become milky exposing crystallization. It is known [4] that at such temperatures in in presence of alkali metal ions in silica glass promotes phase changes to cristobalite. Furthermore, high temperature electrolysis process of silica glass containing sodium with pure gold at negative electrode at 850 K was performed as it was done for natural quartz crystal above. The initial level of current was also small (50 µA), and it was falling down in time of 40 minutes to 1 µA. The same experiments were performed with Ag and Cu introduction in silica glass containing sodium. The spectral-kinetic properties of such samples are presented in Fig.8. The characteristic PL bands of 3O-Si-$O^-$-$Cu^+$ and 3O-Si-$O^-$-$Ag^+$ luminescence centers [2] are well-visible for silica glasses doped with Cu and Ag. The PL decay kinetics show non-exponential behaviour, because of the disordered structure of glass. However, it is known that $Cu^+$ and $Ag^+$ has different luminesce lifetimes in the case of crystal sample. In all cases, the luminescence lifetime is determined by intra ions states differently with respect to 3O-Si-O$^-$- Na$^+$ center were duration of luminescence is determined by charge transfer transition between non-bridging oxygen and sodium [2]. For the case of sample to which we try to introduce Au, there are no significant differences neither in luminescence band nor in luminescence decay kinetics, when comparing not heat-treated sample and sample subjected to high-temperature electrolysis process in the presence of Au. Therefore, the presence of Au$^+$ related luminescence center for Au-doped natural quartz glass is not observed.

*Search for Au$^+$ luminescence center in literature.*

There is still quite complicated picture of Au$^+$ luminescence centers in quartz glass and their possible incorporation in crystal structure. There are only two articles provided different PL bands at 3.6 eV (360 nm) [5] and 2.31 (536 nm) eV [6] for KCl crystals. In both cases the decay kinetics was not measured. Most of the published data were devoted to the luminescence of Ag and Cu in alkali halides, but no data about the Au luminescence. Apparently, the authors are not sure of the existence of a luminescence center associated with Au in alkali metal halides. This is in resonance with our obtained conclusion about lack of such center neither in alkali halides nor in quartz.

**Conclusions**

1) Quartz samples which initially did not contain silver, heat-treated at 1200°C in O$_2$ flowa atmosphere with the presence of metallic Ag, show characteristic Ag$^+$ luminescence band with a maximum at 260 nm. The source of Ag is presumably the remains of metallic Ag on the surface of another sample subjected to high-temperature electrolysis from metallic Ag electrodes.

2) Samples of crystalline quartz, intentionally not activated, were subjected to heat-treatment at 1200°C in O$_2$ atmosphere together with small Au wires, likely containing parasitic impurities of Ag and Cu. Characteristic Ag$^+$ and Cu$^+$ luminescence centers were identified for these samples, and not any gold-related luminescence centers were obtained. Unfortunately, the silica glass tube of the reactor also contains some Cu impurities, so the use of pure (99.99) Au wire did not help to incorporate Au in crystalline quartz and further to identify Au luminescence centers against the background of Cu luminescence.

3) It is assumed that the process of introducing $Ag^+$ (or $Cu^+$) during heat-treatment in $O_2$ flow is due to the fact that Cu and/or Ag react at high temperatures to form an oxygen complex, which can diffuse into the samples and modify existing aluminum-alkali metal centers by replacing the alkali ion ($Me^+$) to $Ag^+$ (or $Cu^+$).

4) It was assumed to be the similar in the case of Au incorporation, but it was entirely different. High temperature electrolysis process using pure Au on negative electron, did not promote Au incorporation and Au oxygen complex cration in natural α-quartz containing aluminum accompanied with alkali and in silica glass containing sodium.

5) Quartz samples heat-treated at 1200°C in $O_2$, which initially did not contain silver, exhibit a characteristic luminescence of the $Ag^+$ center with a wavelength of 260 nm. The source of Ag is presumably the remains of metallic Ag on the surface of another sample subjected to high-temperature electrolysis in metallic Ag electrodes.

6) Samples of crystalline quartz, intentionally not activated, were processed at 1200 ° C in oxygen and intentionally added small gold wires, likely containing parasitic impurities of silver and copper. In these new samples, we only obtained $Ag^+$ and $Cu^+$ luminescence centers, but were unable to identify the luminescence of gold-related centers. Unfortunately, the quartz glass tube also contains copper, so the use of pure (99.99) gold did not help to identify gold luminescence centers against the background of copper luminescence.

7) It is assumed that the process of introducing $Ag^+$ (or $Cu^+$) during oxygen treatment is due to the fact that copper and or silver react at high temperatures to form an oxygen complex, which can diffuse into the samples and modify existing aluminum-alkali metal centers by replacing the alkali ion ($Me^+$) to $Ag^+$ (or $Cu^+$).

**Acknowledgment**

This work was supported by the Latvian Science Council project lzp-2021/1-0215. ML thanks to the support from MikroTik patron, a donation is administrated by the University of Latvia Foundation. Institute of Solid State Physics, University of Latvia as the Center of Excellence has received funding from the European Union's Horizon 2020 Framework Programme H2020-WIDESPREAD-01-2016-2017-TeamingPhase2 under grant agreement No. 739508, project CAMART[2].

**Reference:**

[1] A. Trukhin, "Oxygen treatment effect on luminescence of copper-doped α-quartz crystal," in The 14th International Conference on $SiO_2$, Dielectrics and Related Devices, 2023, p. 6.

[2] A. Trukhin, Silicon Dioxide and the Luminescence of Related Materials: Crystal Polymorphism and the Glass State. Cambridge Scholars Publishing, 2021.

[3] G. Hitherington, K. Jack, and M. Ramsay, "The high-temperature electrolysis of vitreous silica. 1. Oxidation, ultraviolet indused fluorescence, and irradiation color.," Phys Chem Glas, no. 6, pp. 6–15, 1965.

[4] A. G. Boganov et al., "Hydroxyl-Free Quartz Glass for Low-Loss Fiber Optical Waveguides and Its Comparative Radiation-Optical Properties.," Sov J Quantum Electron, vol. 7, no. 5, pp. 558–562, 1977, doi: 10.1070/QE1977v007n05ABEH012547.

[5] L. Bosi, A. Tavecchio, and M. Zelada, "On the KCl:Au+ fluorescence," Solid State Commun., vol. 108, no. 12, pp. 903–905, Nov. 1998, doi: 10.1016/S0038-1098(98)00488-8.

[6] N. E. Lushchik and C. B. Lushchik, "Spectroscopy of the Luminescence Centers in Alkali Halide Crystals Activated by Homologous Series of Ions," Opt. Spectrosc., no. 8, p. 441, 1960.

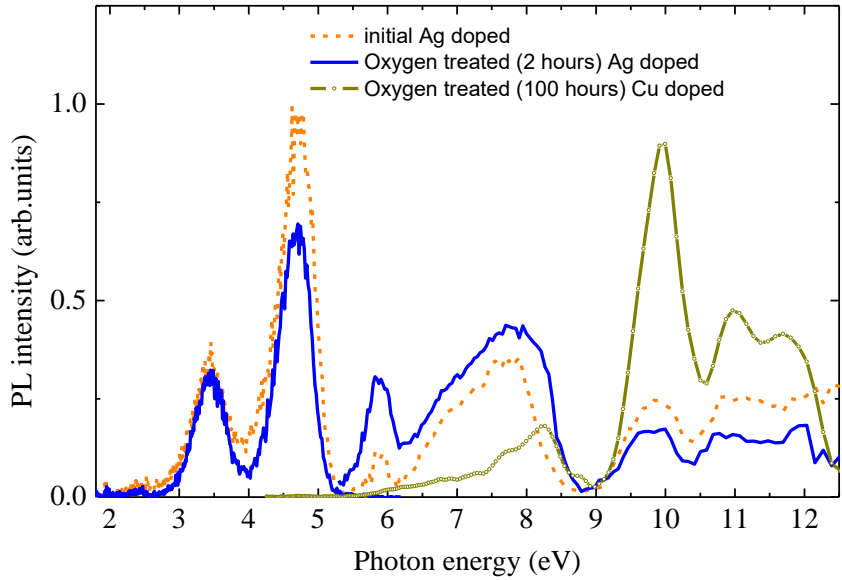

Fig.1 PL spectra of Ag doped α-quartz using electrolysis process, heat-treated and non-treated in $O_2$, and excited with the pulses of $F_2$ laser at 293 K. Both bands belong to $Ag^+$ centers in different structures [2]. The PLE spectra are situated above 5 eV and correspond to treated and non-treated part of Ag doped quart (detected through UFS1 filter comprising both PL band of Ag). The PLE spectrum of quartz doped with Cu and heat-treated for 100 hours is showed for comparison.

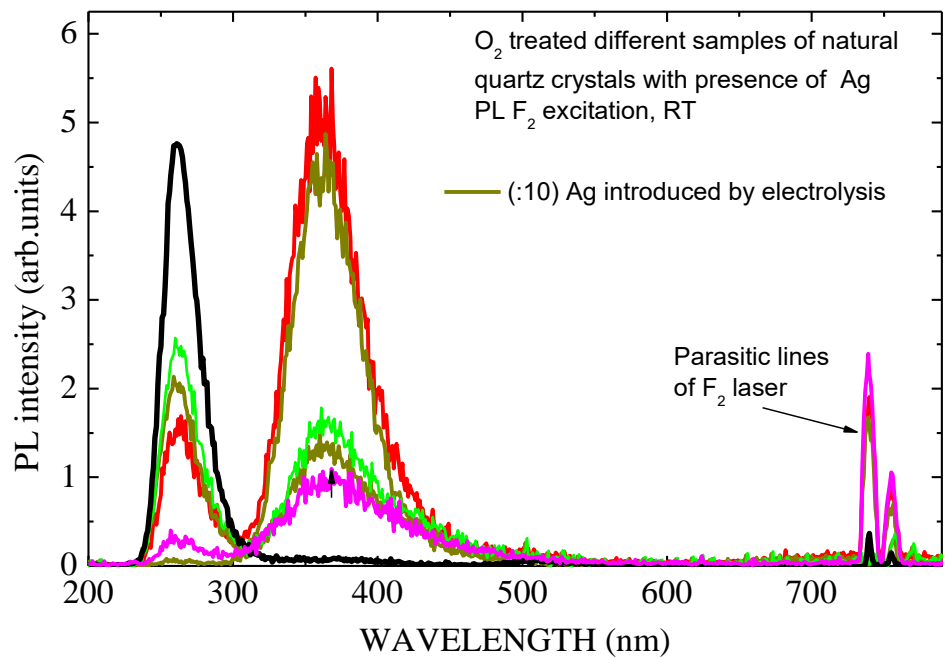

Fig.2 UV PL spectra of different kind of natural quartz crystal samples heat-treated in $O_2$ atmosphere with the presence of Ag.

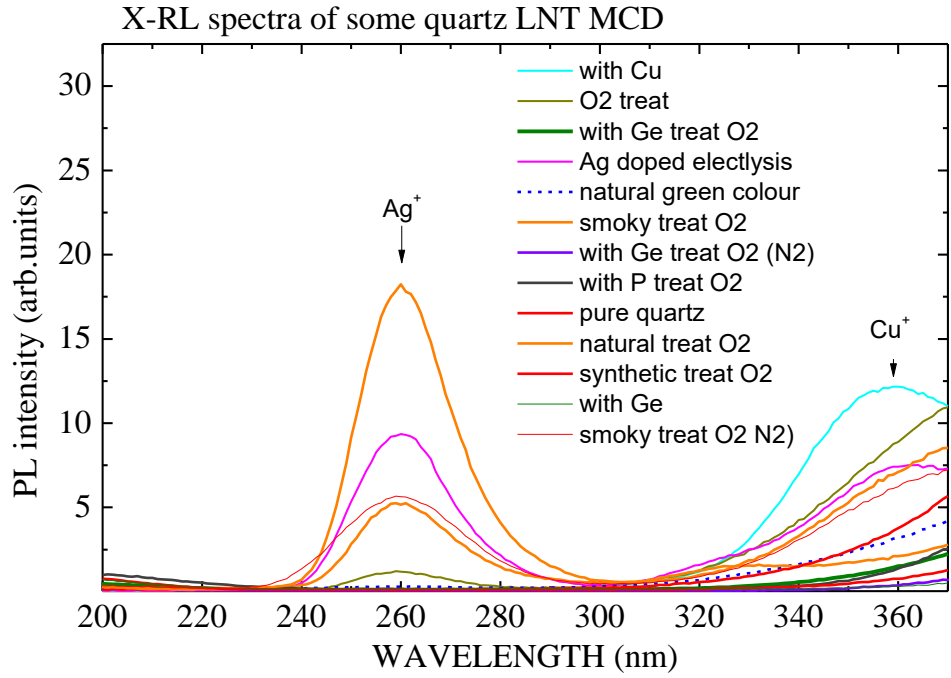

Fig.3 XRL spectra of different kind Quartz samples

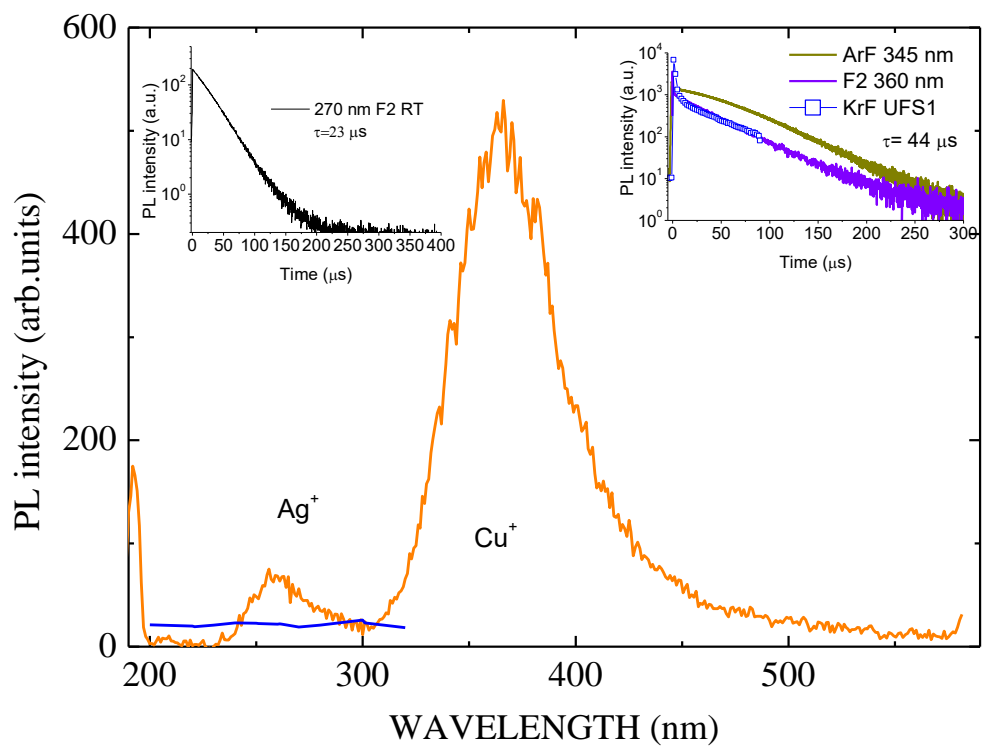

Fig.4 Spectral, decay kinetics of PL parameter of natural crystalline quartz kept at 1200°C in $O_2$ in presence of a Au wire.

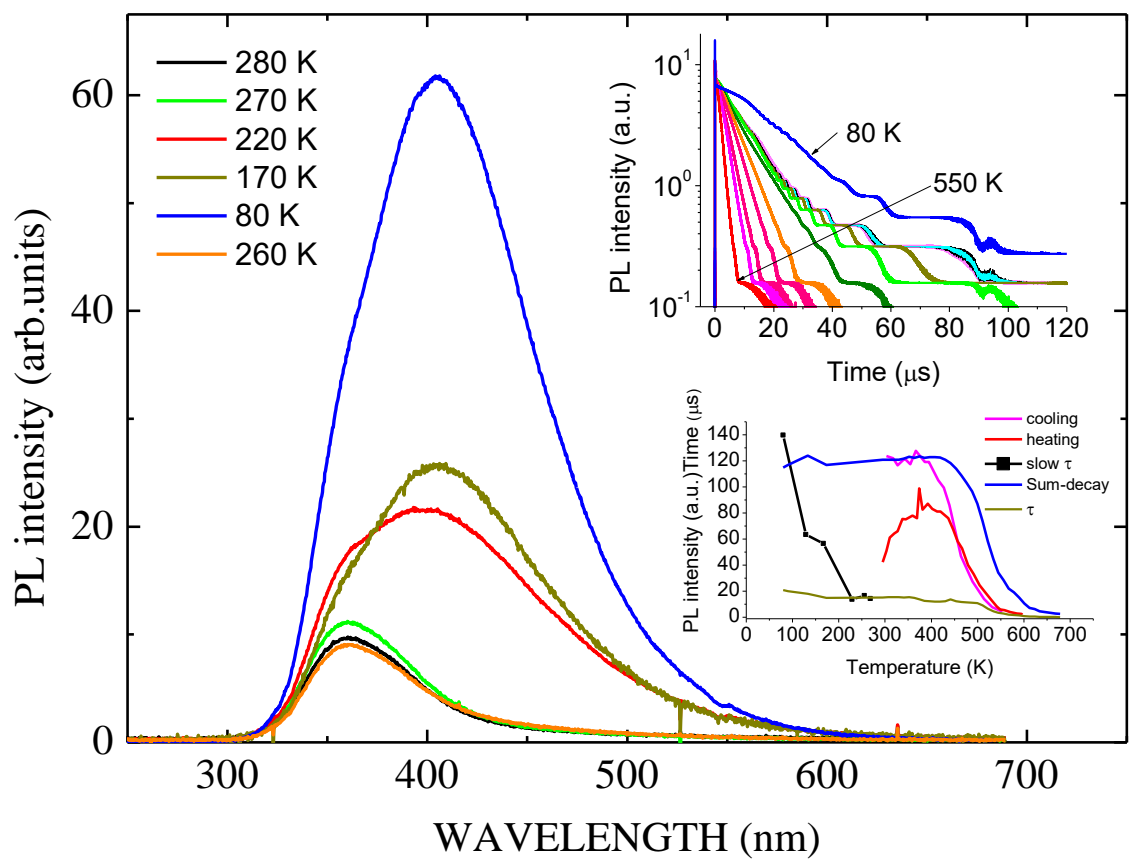

Fig. 5. Spectral, decay kinetics and temperature dependences of PL parameter of natural crystalline quartz kept at 1200°C in $O_2$ in presence of a Au wire.

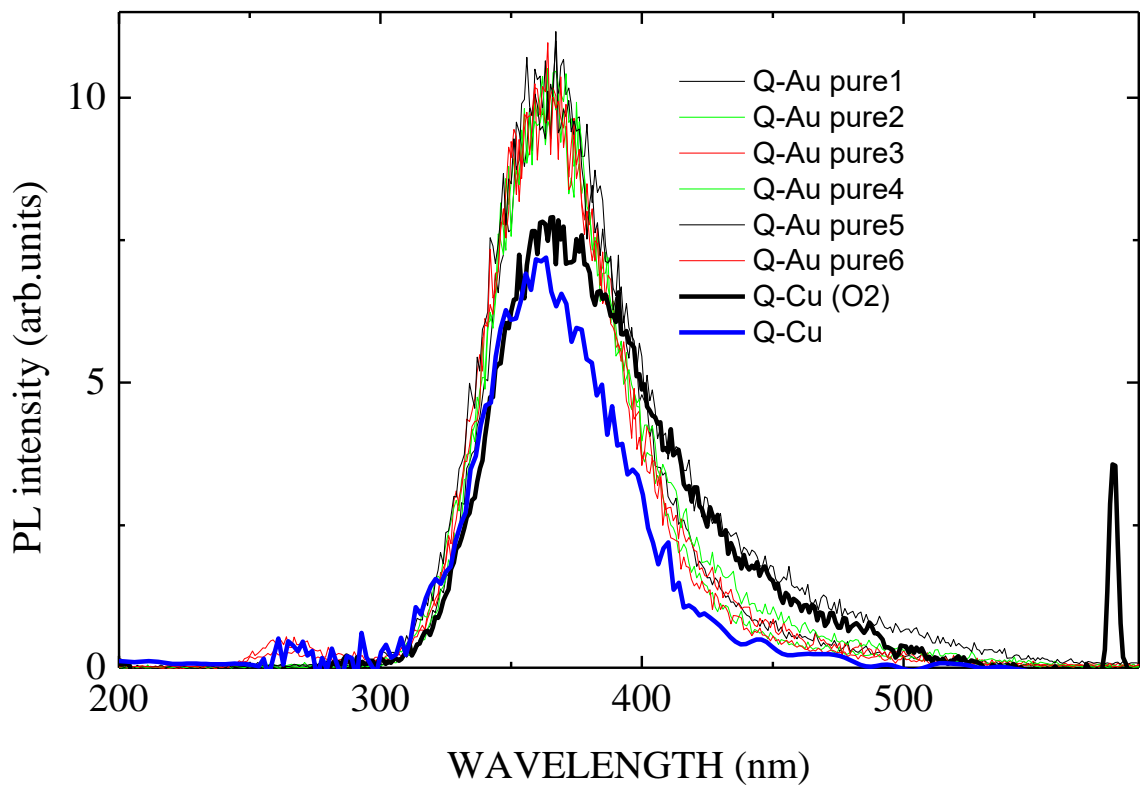

Fig.6. PL spectra of quartz samples heat-treated in $O_2$ atmosphere using metallic Au and quartz samples doped with Cu

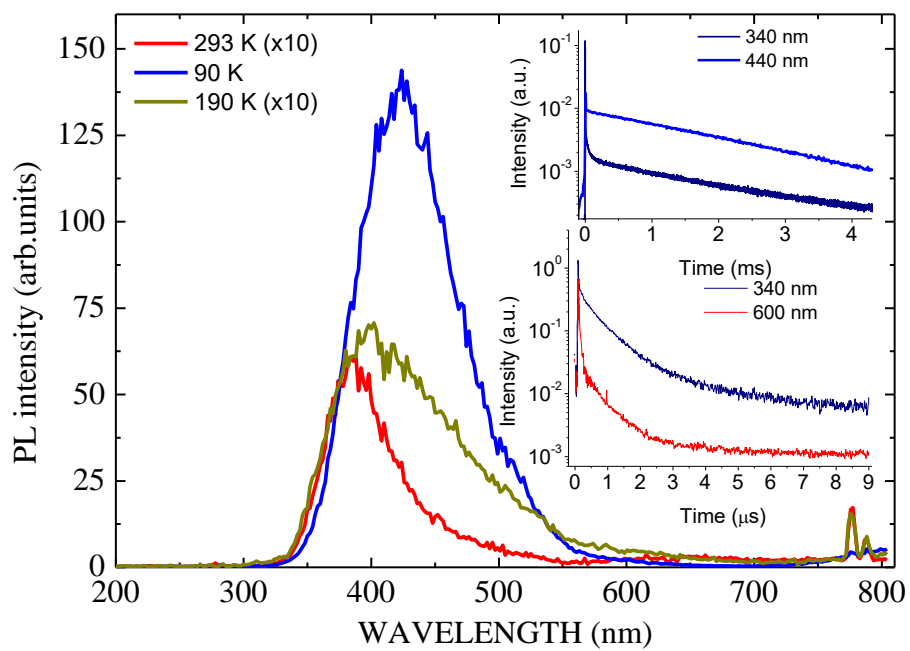

Fig.7 PL spectra and decay kinetics excited with $F_2$ excimer laser of α-quartz samples subjected to electrolysis at 850°C using high purity metallic Au. The decay unequivocally show on practically not changed $[AlO^-_4\text{-}(alkali)^+]$ PL center [2].

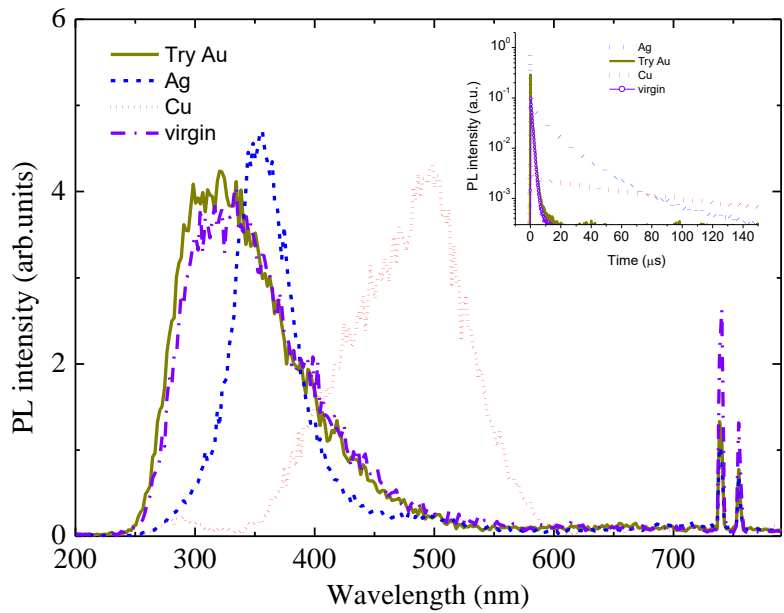

Fig. 8. PL properties of silica glass containing sodium and affected to high temperature electrolysis with Cu, Ag and pure Au electrode. The non-treated sample spectrum is presented as reference.